\documentclass[11pt]{article}
\usepackage{moriond,epsfig}
\usepackage{amsmath}
\usepackage{bbold}
\usepackage{wasysym}
\usepackage{hyperref}
\usepackage{graphics}
\usepackage{url}
\usepackage{graphicx}

\newcommand{\nc}{\newcommand}
\nc{\beq}{\begin{equation}}
\nc{\eeq}{\end{equation}}
\nc{\bea}{\begin{eqnarray}}
\nc{\eea}{\end{eqnarray}}
\nc{\nn}{\nonumber}
\nc{\bi}{\begin{itemize}} 
\nc{\ei}{\end{itemize}}
\nc{\ib}{\item [$\bullet$]}
\nc{\cd}{\cdot}
\nc{\cds}{\cdots}
\nc{\pr}{\prime}
\nc{\tz}{\tilde{z}}
\nc{\msbar}{\overline {\textrm{MS}}}

\nc{\veps}{\varepsilon}
\nc{\as}{\alpha_s}
\nc{\med}{\medskip}

\def\ie{{\sl i.e.}~}

\def\Fig{Fig.~}

\def\eq{Eq.~}

\newbox\charbox
\newbox\slabox
\def\s#1{{      % Feynman slash
        \setbox\charbox=\hbox{$#1$}
        \setbox\slabox=\hbox{$/$}
        \dimen\charbox=\ht\slabox
        \advance\dimen\charbox by -\dp\slabox
        \advance\dimen\charbox by -\ht\charbox
        \advance\dimen\charbox by \dp\charbox
        \divide\dimen\charbox by 2
        \raise-\dimen\charbox\hbox to \wd\charbox{\hss/\hss}
        \llap{$#1$}
}}
\newlength{\nseparation}
\nc{\lag}{\cal L }
\nc{\matx}{\left|\cal {M}\right|^2}
\nc{\lqcd}{\Lambda_\textrm{QCD}}
\nc{\really}{\stackrel{!}{=}}

\newcommand{\eqn}{equation}

\newcommand{\lb}{\left(}
\newcommand{\rb}{\right)}

\newcommand{\al}{\alpha}
\newcommand{\M}{\mathcal{M}}

\newcommand{\D}{\mathcal{D}}

\newcommand{\ph}{\hat{p}}

\begin{document}
\vspace*{4cm}
\title{AN ALTERNATIVE SUBTRACTION SCHEME FOR NLO CALCULATIONS}

\author{T. ROBENS}

\address{School of Physics and Astronomy, University of Glasgow, Glasgow, G12 8QQ, Scotland, UK}

\author{C.H. CHUNG and M. KR\"AMER}

\address{Institute for Theoretical Particle Physics and Cosmology, 
RWTH Aachen,
52056 Aachen, Germany}

\maketitle\abstracts{We present a new subtraction scheme for next-to-leading
order QCD calculations, where the momentum mapping and the splitting functions have been derived in the context of an improved
parton shower formulation. A main advantage of our scheme is the significantly reduced number of momentum mappings in the subtraction terms compared to standard schemes.
We present the major features of our scheme and discuss the process $e\,q\,\rightarrow\,e\,q\,(g)$
 in more detail.}

\section{Introduction}
Both the further validation of the Standard Model (SM) as well as searches for new physics beyond the SM require an exact knowledge of the SM signals at at least Next-to-leading order (NLO). For precise differential predictions, these NLO corrections need to be included in Monte Carlo Event Generators.
However, an increase of final state particle multiplicity in the LO process in such codes directly translates to an increase of the computational runtime.
This is partially caused by the treatment of infrared (IR) singularities: For standard subtraction schemes, the number of momentum mappings and Born matrix reevaluations rapidly increases with the number of final state particles. We here present a new scheme which significantly reduces the number of the momentum mappings in the real emission subtraction terms.
\section{Subtraction schemes}%: general setup}
We consider a generic jet cross-section $\sigma$ with
{
\begin{eqnarray}
\sigma &= & \sigma^{\text{LO}}+\sigma^{\text{NLO}}
\, = \, \int_m d\sigma^B + \int_m d\sigma^V + \int_{m+1}d\sigma^R\,,
\end{eqnarray}}
\noindent
where $\sigma^{\text{B}}$, $\sigma^{\text{V}}$, and
$\sigma^{\text{R}}$ denote the LO, virtual and real-emission
contributions, and with $m\,(m+1)$ partons in the final state in the LO (real emission) phase space. 
The IR poles, which are inherent in both $d\sigma^V$ and $d\sigma^R$, cancel in the sum of these two terms; however, the individual pieces are
divergent and can thus not be integrated numerically.
Subtraction schemes resolve this issue by introducing local counterterms, which match
the behaviour of the real-emission matrix element in each singular
 region. Subtracting the counterterms from the
real-emission matrix elements and adding back the corresponding
one-particle integrated counterparts to the virtual contribution then
results in overall finite integrands
{
%\small
\bea
\label{countertermfinite85}
&&\sigma^{\text{NLO}}\,=\,
\underset{\textrm {finite} }
{\underbrace{\int_{m}\,d\sigma^V+\int_{m+1}\,d\sigma^A}}   +
\underset{ \textrm {finite} }
{\underbrace{\int_{m+1}\left[ d\sigma^R-d\sigma^A\right]}}.
\eea
}
\noindent
The construction of the local counterterms, collectively denoted by
$d\sigma^A$ in \eq (\ref{countertermfinite85}), relies on the
factorisation of the real-emission matrix element in the singular (\ie
soft and collinear) limits:
{
%\small
${\cal M}_{m+1}(\{\hat p\}_{m+1}) \longrightarrow \sum_{\ell}
v_{\ell}(\{\hat p\}_{m+1})\otimes {\cal M}_m(\{p\}_m)\,,$ 
}
where {\small ${\cal M}_m\,\lb {\cal M}_{m+1}\rb$} and {\small ${\cal M}_m$} denote {\small $m\,(m+1)$}
matrix elements and the $v_{\ell}$ are generalised
splitting functions containing the complete singularity structure. 
As ${\cal M}_{m+1}$ and ${\cal M}_{m}$
live in different phase spaces,
a mapping $\{\hat p\}_{m+1} \to \{p\}_m$ needs to be introduced, which conserves four-momentum and guarantees onshellness
for all external particles in both phase spaces.
Squaring and averaging over the splitting functions 
then leads to subtraction terms of the form
{
\begin{\eqn}\label{eq:wlkdef}
W_{\ell\,k}\,=\,v_\ell(\{\hat p, \hat f\}_{m+1}, \hat s_j, \hat s_\ell, s_\ell)\,
   v_k(\{\hat p, \hat f\}_{m+1}, \hat s_j, \hat s_k,    s_k)^* \,
    \delta_{\hat s_\ell, s_\ell} \,
    \delta_{\hat s_k, s_k}\,,
\end{\eqn}
}
where our notation follows $p_\ell\,\rightarrow\,\ph_\ell\,\ph_j$ for parton splitting, and $f$ denotes the parton flavour. We now distinguish two different kinds of subtraction terms: 1) direct squares where $k\,=\,\ell$, which contain both collinear and soft singularities; 2) soft interference terms, where $k\,\neq\,\ell$. The latter contain only soft singularities and vanish if $f_j\,\neq\,g$; here, $v$ is replaced by the eikonal approximation of the splitting function $v^{\textrm {eik}}$. These terms explicitly depend on the spectator four momentum $\ph_k$.
In the following, we symbolically  write ${\cal D}_{\ell}$ for terms of the form $W_{\ell,\ell}$ and $W_{\ell,k}$. We then have
$d\sigma^A = \sum_{\ell} {\cal D}_{\ell} \otimes d\sigma^B\,,$
with $\otimes$ representing phase-space, spin and colour convolutions. Integrating the subtraction term $d\sigma^A$ over the one-parton
unresolved phase space, $d\xi_p$, yields an infrared- and
collinear-singular contribution ${\cal V}_{\ell}\,=\,\int\,d\xi_p {\cal D}_{\ell}$
which needs to be combined with the virtual cross section to yield a
finite NLO cross section
{
%\small
\beq
\label{eq:nlofin}
\sigma^{\text{NLO}} =
\int_{m}\Big[ d\sigma^V+\sum_{\ell}{\cal V}_{\ell}\otimes d\sigma^B\Big]  +
\int_{m+1}\Big[ d\sigma^R-\sum_{\ell}{\cal D}_{\ell}\otimes d\sigma^B\Big]\,.
\eeq
}
In this form, the NLO cross section can be integrated numerically over
phase space using Monte Carlo methods. 
The jet
cross-section $\sigma$ has to be defined in a infrared-safe way by the
inclusion of a jet-function {\small $F_J$}, which satisfies {\small $F_J^{(m+1)}\,\to
\,F_J^{(m)}$} in the collinear and infrared limits.
\subsection{Major features of new subtraction scheme}
Our scheme\cite{Robens:2010zr} \cite{Chung:2010fx} \cite{chenghanthesis}
uses the splitting functions of an improved parton shower\cite{Nagy:2007ty} \cite{Nagy:2008ns} \cite{Nagy:2008eq}
as the basis for the local subtraction terms. 
The main advantage\footnote{An additional advantage stems from the use of common splitting functions in the shower and the subtraction scheme which facilitates the matching of shower and parton level NLO calculation.} of our scheme is a novel momentum mapping for final state emitters:
for 
$\hat{p}_{\ell} + \hat{p}_{j} \to
p_{\ell}$, we redistribute the momenta according to the global mapping\footnote{The parameter definitions are
$y = \frac{P_\ell^2}{2\, P_\ell\cdot Q - P_\ell^2},\,
a_\ell \,=\, \frac{ Q^2}{2\, P_\ell\cdot Q - P_\ell^2},\, \lambda\,=\,\sqrt{\left(1+y\right)^2-4\,a_\ell\,y}, 
\,P_\ell \,= \, \hat p_\ell+\hat p_j,$\\$ Q\, = \, \hat p_a + \hat p_b \,=\,  \sum_{n = 1}^{m+1}\, \hat p_n .
$
 The Lorentztransformation 
$
\Lambda(K,\hat K)^{\mu}_{\;\;\nu} $
 is a function of $K\,=\,Q-p_\ell,\,\hat K\,=\,Q-P_\ell$.}
{
%\small
\begin{eqnarray}
\label{eq:fin_map}
&&p_\ell\,=\,\frac{1}{\lambda}\,(\hat p_\ell + \hat p_j)-\frac{1 - \lambda + y}{2\, \lambda\, a_\ell}\, Q,\quad
  p_n^\mu\,=\,\Lambda (K,\hat{K})^\mu{}_\nu \,\hat p^{\nu}_{n} ,\quad n\notin\{\ell,a,b\}\,,
\end{eqnarray}
}
where $n$ labels all partons in the $m$ particle phase space which do not participate in the inverse splitting. 
We here consider the resulting implications on a purely gluonic process with only $g(p_{\ell})\,\rightarrow\,g(\ph_{\ell})\,g(\ph_{j})$ splittings. For final state emitters, the real emission subtraction terms are %then given by
{
\begin{eqnarray}\label{eq:finsum}
d\sigma^{A,p_\ell}_{ab}(\hat{p}_a,\hat{p}_b)&=&\frac{N_{m+1}}{\Phi_{m+1}}
\sum_{i=j}\,\D_{ggg}(\ph_{i})\,|\M_{\text{Born},g}|^{2}(p_{a},p_{b};p_{\ell},p_{n}),
\end{eqnarray}
}
where {\small $\M_{\text{Born},g}$}
is the underlying Born matrix element for the process
{\small $p_{a}+p_{b}\,\rightarrow\, \sum_n p_{n}+p_{\ell} $}, $\Phi,\,N$ are flux and combinatoric factors,
and the sum $i\,=\,j$ goes over all $i$ final state gluons. $\D_{ggg}$ contains both collinear
and soft interference terms:
{
%\small
\begin{\eqn}
\D_{ggg}(\ph_{i})\,=\,\D^{\text{coll}}_{ggg}(\ph_{i})\,+\sum_{k=a,b}\,\D^{\text{if}}(\ph_{i},\hat{p}_{k})\,+\,\sum_{k\,\neq\,i}\D^{\text{if}}(\ph_{i},\ph_{k}).
\end{\eqn}
}
While the collinear subtraction terms only depend on the four-momenta $\ph_{j},\,\ph_{\ell}$, the soft interference terms have an additional dependence on the four-momentum of the spectator $\ph_{k}$ (cf Eq. (\ref{eq:wlkdef})). The sums over $(a,b)$ and $(k\,\neq\,i)$ then run over all possible spectators in soft interference terms; however, {\sl there is a unique mapping per combination $(\hat{\ell},\hat{i})$ in Eq. (\ref{eq:finsum}) which is independent of the spectator momenta $\ph_{k}$ in the soft interference terms.} Therefore, the underlying matrix element $|\M_{\text{Born},g}|^{2}(p_{a},p_{b};p_{\ell},p_{n})$ only needs to be evaluated once for all possible spectators. {\sl This is indeed the main feature of our scheme}, which,  for $N$ parton final states, leads to a scaling behaviour $\sim N^2/2$
for the number of momentum mappings and LO matrix element reevaluations in the real emission subtraction terms.
\section{Results for $e\,q\,\rightarrow\,e\,q\,(g)$}
\begin{figure}[t]
%\begin{minipage}{0.49\textwidth}
\centering
\includegraphics[width=0.4\textwidth]{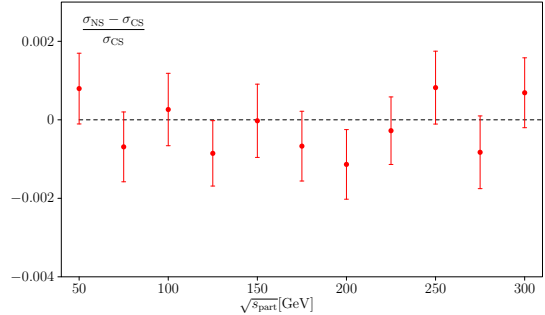}
\vspace{4mm}
\caption{\label{fig:dis_comp} Partonic cross sections for DIS subprocess $e q\,\rightarrow\,e q
  (g)$, as a function of parton level (HERA-like) cm energies, with
  angular cuts $\cos\,\theta_{ee}<0.8$.
% Partonic cross section prior
 % to convolution with PDFs.  
Relative difference between NLO contributions using Nagy-Soper (NS) and Catani Seymour (CS) subtraction terms. Errors are integration errors; results agree at the permil-level.}
\end{figure}

The improved scaling behaviour of the scheme leads to more complicated integrated subtraction terms, which partially need to be evaluated numerically. As an example, we discuss the integrated subtraction terms and the resulting integrand in the two particle phase space for the DIS sub-process
$e(p_{i})\,+\,q(p_{1})\,\rightarrow\,e(p_{o}) q(p_4)\,\lb g(p_{3}) \rb$
on parton level.
The effective two particle phase space contribution is given by
{%\small
\begin{eqnarray*}
\lefteqn{|\M|_\text{Born}^{2}(2\,p_{i}\cdot p_1)\,+\,2\,\text{Re}\lb\M_\text{Born}\,\M^*_\text{virt} \rb(2\,p_{i}\cdot p_1) \,+\,\sum_\ell \int\,d\xi_{p}\D_\ell\,|\M|^2_\text{Born}(2 p_i\,\cdot x p_1)=}\nn\\
&&\int^{1}_{0}\,dx\,\Bigg\{\,\frac{\al_{s}}{2\,\pi}\,C_{F}\,\delta(1-x)\,\left[-9\,+\,\frac{1}{3}\pi^{2}\,-\,\frac{1}{2}\text{Li}_{2}[(1-\tz_{0})^{2}]\,\nn\right.
\left.+\,2\,\ln\,2\,\ln\tz_{0}\,+\,3\,\ln\tz_{0}\,+\,3\,\text{Li}_{2}(1-\tz_{0})\nn \right. \\
&&\left.\,+\,I^\text{tot,0}_\text{fin}(\tz_{0})\,+\,I^{\text 1
}_\text{fin}(\tilde{a}) \right. \bigg]
\,+\,K_\text{fin}^\text{tot}(x;\tilde{z})\,+\,P^\text{tot}_\text{fin}(x;\mu^{2}_{F})  \Bigg\}|\M|^{2}_\text{Born}(2 p_i \cdot x\,p_{1}),
\end{eqnarray*}
}
where 
{%\small
\begin{eqnarray}\label{eq:dis_kp}
K^\text{tot}_\text{fin}(x;\tz)&=&\frac{\al_{s}}{2\,\pi}\,C_{F}  \Bigg\{\frac{1}{x}\left[
  2\,(1-x)\,\ln\,(1-x)\,-\,\lb\frac{1+x^{2}}{1-x}\rb_{+}\,\ln\,x\nn\right.
 \left. \,+\,4\,x\,\lb\frac{\ln(1-x)}{1-x}\rb_{+}\right]\,+\,I_\text{fin}^{1}(\tilde{z},x)\Bigg\}
\end{eqnarray}
}
The integrals {\small $I^\text{tot,0}_\text{fin}(\tz_{0}),\,I^{\text 1
}_\text{fin}(\tilde{a}),I_\text{fin}^{1}(\tilde{z},x)$} all need to be integrated numerically; as an example we give
{%\small
\begin{\eqn}
{ I^{1}_\text{fin}(\tilde{z},x)}\,=\,\frac{2}{(1-x)_{+}}\,\,\frac{1}{\pi}\int^{1}_{0}\,\frac{dy'}{y'}\,
\left[\int^{1}_{0}\,\frac{dv}{\sqrt{v\,(1-v)}}\,{ \frac{\tilde{z}}{{
  N(x,y',\tilde{z},v)}}}-1\right]
\end{\eqn}
}
which explicitly depends on the real emission four vectors $\ph_3,\,\ph_4$ through the variables  {\small $N, \tilde{z} $} \footnote{Exact definitions are ${ N}\,=\,\frac{ \ph_{3} \cdot \ph_{4}}{\ph_{4}\cdot
  \hat{Q}}\,\frac{1}{1-x}\,+\,y',\tilde{z}\,=\,\frac{1}{x}\frac{p_{1} \cdot
   \ph_{4}}{ \ph_{4} \cdot \hat{Q}},\,\tilde{z}_{0}\,\stackrel{y'\,\rightarrow\,0}{=}\tilde{z}$.}.
 Note that $\ph_3$ needs to be reconstructed in the two particle phase space; 
$\ph_4$ can be obtained from the inverse of Eq.~(\ref{eq:fin_map}).
We perform the numerical evaluation of all integrals in parallel to the phase space integration; however, the integrals are process-independent expressions with a functional dependence on maximally two input parameters and can therefore be evaluated generically such that no additional numerical evaluation is necessary in future implementations of our scheme. 
We have compared the results for the parton level process $q\,e\,\rightarrow\,q\,e\,(g)$ numerically with an implementation
of the Catani-Seymour dipole subtraction\cite{Catani:1996vz}.  \Fig \ref{fig:dis_comp} shows the
behaviour of the differences
between the application of the two schemes at parton level for varying
(HERA-like) center-of-mass energies, where we applied an appropriate jet function to guarantee IR safety of {\small $d\sigma^\text{B}$}.
The results agree on the
per-mil level, therefore validating
our subtraction prescription. We equally observed that this cancellation is non-trivial as the contributions from
different phase space integrations vary widely in magnitude for the
two schemes.
 %\begin{figure}
%\begin{minipage}{0.49\textwidth}
%\centering
%\includegraphics[width=0.8\textwidth, angle = -90]{0.1.ps}
%\end{figure}
%\end{minipage}
%\hspace{0.5cm}
%\begin{minipage}{0.49\textwidth}
%\begin{figure}
%\centering
%\includegraphics[width=0.8\textwidth, angle = -90]{0.05.ps}
%\end{minipage}
%\centering
%\includegraphics[width=0.4\textwidth, angle = -90]{0.2.ps}
%\caption{\label{fig:dis_comp} Comparison of Catani Seymour and Nagy Soper subtraction terms for DIS sub process $e q\,\rightarrow\,e q (g)$, varied parton level cm energies, with angular cuts $\cos\,\theta_{ee}<0.95$ and $\cos\theta_{ee}<0.9$ respectively. Results agree on the permil level. }
%\begin{center}
%\begin{minipage}{0.45\textwidth}
%\centering
%\includegraphics[width=0.8\textwidth, angle = -90]{0.2.tot.ps}
%\end{figure}
%\end{minipage}
%\hspace{0.4cm}
%\begin{minipage}{0.45\textwidth}
%\begin{figure}
%\centering
%\includegraphics[width=0.8\textwidth, angle = -90]{0.2.rel.ps}
%\includegraphics[width=\textwidth]{0.2.rel.eps}
%\end{minipage}
%\caption{\label{fig:dis_comp2} Comparison of Catani Seymour and Nagy Soper subtraction terms for DIS sub process $e q\,\rightarrow\,e q (g)$, varied parton level (HERA-like) cm energies, with angular cuts $\cos\,\theta_{ee}<0.8$. Left: NLO and LO total results for both schemes; relative NLO corrections are around $3-3.5\%$. Right: relative difference between NS and CS results. Partonic cross section prior to convolution with PDFs. Results agree on the permil level.}
%\end{center}
%\end{figure}
%\begin{figure}[t]
%\begin{minipage}{0.49\textwidth}
%\centering
%\includegraphics[width=0.45\textwidth, angle = -90]{0.2.diff.ps}
%\includegraphics[width=0.4\textwidth]{0.2.diff.eps}
%\vspace{4mm}
%\end{figure}
%\end{minipage}
%\hspace{0.5cm}
%\begin{minipage}{0.49\textwidth}
%\begin{figure}
%\centering
%\includegraphics[width=0.8\textwidth, angle = -90]{0.05.ps}
%\end{minipage}
%\centering
%\includegraphics[width=0.4\textwidth, angle = -90]{0.2.ps}
%\caption{\label{fig:dis_comp_diff} As Figure \ref{fig:dis_comp}. Behaviour of the {\sl difference} $\Delta^{(2,3)}_{\text{NS,CS}}\,=\,\sigma^{(2,3)}_{\text{NS}}-\sigma^{(2,3)}_{\text{CS}}$ between the two schemes for the two particle final state phase space (green, dashed) and three particle final state phase space (blue, solid), respectively. In the sum $\Delta^{(2)}_{\rm NS,CS}+\Delta^{(3)}_{\rm NS,CS}$ (black dots), the large differences cancel. In the new scheme, subtractions have been restricted to singular regions.
%}
%\end{figure}

\section{Conclusions and Outlook}
Our current results present a first step in the establishment of a new subtraction scheme. The scheme we propose reduces the number of momentum mappings and therefore of reevaluation calls of the underlying LO matrix element. We have derived the splitting functions and validated our scheme by reproducing the literature results for various $1\,\rightarrow\,2$ and $2\,\rightarrow\,2$ processes\cite{Robens:2010zr} \cite{Chung:2010fx} \cite{chenghanthesis}.
The application to processes with more than two particles in the final state is nearly finished\cite{us2}; similarly, an implementation of the scheme within the Helac framework is underway \cite{NS-helac}. 
An additional advantage of the scheme comes from the use of splitting functions which are derived from an improved parton shower; this approach promises advantages for the combination of NLO parton level calculations and parton showers.
{%\small
\section*{Acknowledgements}TR wants to especially thank Zoltan Nagy, Dave Soper, and Zoltan Trocsanyi for input and many valuable discussions, as well as the organizers of the Moriond 2011 QCD session for financial support and perfect weather during the midday breaks. This research was partially supported by the DFG SFB/TR9, the DFG Graduiertenkolleg
``Elementary Particle Physics at the TeV Scale'', the Helmholtz
Alliance ``Physics at the Terascale'', the BMBF, the STFC and the EU
Network MRTN-CT-2006-035505.
}
\section*{References}
\bibliography{NS_subtraction}

\begin{thebibliography}{1}

\bibitem{Robens:2010zr}
T.~Robens and C.~H. Chung.
\newblock {Alternative subtraction scheme using Nagy Soper dipoles}.
\newblock {\em PoS}, RADCOR2009:072, 2010.

\bibitem{Chung:2010fx}
Cheng-Han Chung, Michael Kramer, and Tania Robens.
\newblock {An alternative subtraction scheme for next-to-leading order QCD
  calculations}.
\newblock 2010.

\bibitem{chenghanthesis}
C.H.Chung.
\newblock {\em An alternative subtraction scheme using Nagy Soper dipoles}.
\newblock PhD thesis, RWTH Aachen University, Aachen, Germany, 2011.

\bibitem{Nagy:2007ty}
Zoltan Nagy and Davison~E. Soper.
\newblock Parton showers with quantum interference.
\newblock {\em JHEP}, 09:114, 2007.

\bibitem{Nagy:2008ns}
Zoltan Nagy and Davison~E. Soper.
\newblock {Parton showers with quantum interference: leading color, spin
  averaged}.
\newblock {\em JHEP}, 03:030, 2008.

\bibitem{Nagy:2008eq}
Zoltan Nagy and Davison~E. Soper.
\newblock {Parton showers with quantum interference: leading color, with spin}.
\newblock {\em JHEP}, 07:025, 2008.

\bibitem{Catani:1996vz}
S.~Catani and M.~H. Seymour.
\newblock A general algorithm for calculating jet cross sections in NLO QCD.
\newblock {\em Nucl. Phys.}, B485:291--419, 1997.

\bibitem{us2}
Cheng~Han Chung, Michael Kr{\"a}mer, and Tania Robens.
\newblock In preparation.

\bibitem{NS-helac}
Giuseppe Bevilacqua, Michal Czakon, Michael Kr{\"a}mer, and Michael Kubocz.
\newblock Work in progress.

\end{thebibliography}
\end{document}